\documentclass[10pt,twocolumn,letterpaper]{article}

\usepackage[pagenumbers]{wacv} 
\usepackage{tabularx}
\usepackage{adjustbox}
\usepackage{subcaption} 
\usepackage{graphicx}
\usepackage{caption}
\usepackage{cuted}
\usepackage{makecell}
\usepackage{flushend} 
\usepackage[accsupp]{axessibility}
\definecolor{wacvblue}{rgb}{0.21,0.49,0.74}
\usepackage[pagebackref,breaklinks,colorlinks,allcolors=wacvblue]{hyperref}

\def\wacvPaperID{244} 
\def\confName{WACV}
\def\confYear{2026}

\title{Self-Supervised Compression and Artifact Correction for Streaming \\ Underwater Imaging Sonar}

\author{
Rongsheng Qian$^1$, Chi Xu$^1$, Xiaoqiang Ma$^2$, Hao Fang$^1$,
Yili Jin$^{3,1}$, William I. Atlas$^4$, Jiangchuan Liu$^1$\\
$^1$Simon Fraser University, BC, Canada,
$^2$Douglas College, BC, Canada,\\
$^3$McGill University, QC, Canada,
$^4$Wild Salmon Center, OR, USA\\
\tt\small\{rqa4, chix\}@sfu.ca, mxqcs@ieee.org, fanghaof@sfu.ca, yili.jin@mail.mcgill.ca,\\
\vspace{-0.6cm}
\tt\small watlas@wildsalmoncenter.org, jcliu@sfu.ca
}
\begin{document}
\maketitle


\begin{abstract}

Real-time imaging sonar is crucial for underwater monitoring where optical sensing fails, but its use is limited by low uplink bandwidth and severe sonar-specific artifacts (speckle, motion blur, reverberation, acoustic shadows) affecting up to 98\% of frames. We present \textbf{SCOPE}, a self-supervised framework that \emph{jointly} performs compression and artifact correction without clean–noise pairs or synthetic assumptions. SCOPE combines (i) Adaptive Codebook Compression (ACC), which learns frequency-encoded latent representations tailored to imaging sonar, with (ii) Frequency-Aware Multiscale Segmentation (FAMS), which decomposes frames into low-frequency structure and sparse high-frequency dynamics while suppressing rapidly fluctuating artifacts. A hedging training strategy further guides frequency-aware learning using low-pass proxy pairs generated without labels.
Evaluated on months of in-situ ARIS sonar data, SCOPE achieves a structural similarity index (SSIM) of 0.77, representing a 40\% improvement over prior self-supervised denoising baselines, at bitrates down to $\leq 0.0118$ bpp. It reduces uplink bandwidth by more than 80\% while improving downstream detection. The system runs in real time, with 3.1 ms encoding on an embedded GPU and 97 ms full multi-layer decoding on the server end. SCOPE has been deployed for months in three Pacific Northwest rivers to support real-time salmon enumeration and environmental monitoring in the wild. Results demonstrate that learning frequency-structured latents enables practical, low-bitrate sonar streaming with preserved signal details under real-world deployment conditions.

\end{abstract}

\vspace{-0.3cm}

\section{Introduction}

Traditional sonar operated at high power for longer range and wider coverage, but had limited resolution, making it difficult to identify small targets and resolve fine details of objects. In contrast, modern imaging sonars provide high-resolution imagery at lower power, enabling detailed observation of underwater scenes and real-time monitoring in conditions where optical and infrared sensors are unreliable~\cite{jones2021adaptive,wei2022monitoring}. By transmitting acoustic pulses and reconstructing echoes into spatial imagery, sonar allows perception in turbid or low-light environments where cameras fail~\cite{hu2024underwater,shi2024sonar}. Advances in hardware affordability and access to 5G and satellite connectivity have expanded sonar-based analytics in remote settings, including offshore rescue~\cite{hu2024underwater}, subsea infrastructure inspection~\cite{shi2024sonar}, and seasonal fishery management~\cite{10.1145/3666025.3699323,kay2022caltechfishcountingdataset}, highlighting its societal value in safety, industrial support, and ecological monitoring~\cite{jahanbakht2021internet}.

Despite its growing importance, it remains difficult to reliably perform streaming imaging sonar in wild, uncontrolled environments. Limited infrastructure and unstable uplinks restrict transmission capacity~\cite{10228912}, while the imagery itself is degraded by complex artifacts. For instance, in our field studies (see Suppl.), a single satellite uplink (e.g., Starlink) averaged 16.1 Mbps with a standard deviation of 5.7 and fluctuated between 1.1 and 39.9 Mbps. When shared across multiple observation points, the per-site bandwidth dropped to an average of 4.7 Mbps with a standard deviation of 1.9 and a range of 0 to 14 Mbps. This is well below the more than 24 Mbps required to stream compressed sonar video at full resolution (1146x2138@15 FPS), leaving a substantial portion of data untransmitted (see Suppl.). H.264/H.265 at the same CRF reach only $\sim$8.03 Mbps on standard video, showing imaging sonar data is harder to compress (see Suppl.), due to pixel-wise noise fluctuations \cite{Kwan}. Meanwhile, imaging sonar suffers from persistent speckle, motion blur, reverberations, and acoustic shadows~\cite{10274511,rs16152815}. Motion blur affects 98\% of frames~\cite{kay2022caltechfishcountingdataset,jones2021adaptive}, while reverberations and shadows can obscure objects by up to 6× their actual size~\cite{10.1145/3666025.3699323}. These conditions reduce perceptual quality, increase entropy, and hinder both compression efficiency and downstream analysis.

The combined effects of bandwidth limitations and image artifacts show that compression and correction cannot be treated separately. Artifacts increase entropy and reduce coding efficiency, while compression that is not artifact-aware can obscure small or camouflaged targets. Existing research provides only partial answers. Deep learning–based image and video compression has advanced rapidly, with VAE-based models learning compact latent spaces~\cite{10.1145/3664647.3681354,Jia_2024_CVPR,lu2024,Zhang_2024_CVPR,10.1145/3581783.3613799,10.1145/3664647.3681242,Duan_2024_CVPR}, and diffusion or transformer-based approaches achieving strong rate–distortion trade-offs~\cite{10.1145/3664647.3681336,Jiang_2023,Guo_2023,10.1145/3581783.3611960}. Yet these methods are tuned for natural imagery and often fail to preserve the small, sparse motions characteristic of sonar. In parallel, self-supervised denoising approaches, including blind-spot networks~\cite{chihaoui2024, lehtinen2018, chen2024, 8954066} and diffusion models~\cite{Zeng_2024_CVPR, cheng2024}, reduce dependence on clean labels, but typically assume noise is independent or Gaussian. This prevents them from handling environment-dependent distortions such as motion blur, reverberation, and acoustic shadows. These limitations highlight the need for a sonar-specific approach that unifies compression and artifact correction within a single framework.

To address these challenges, we present \textbf{SCOPE}, a real-time framework for \textbf{S}elf-supervised \textbf{CO}mpression and artifact \textbf{CO}rrection in imaging sonar \textbf{P}rocessing and str\textbf{E}aming. SCOPE is built on a Variational Autoencoder (VAE) backbone that performs latent-space compression, but extends it with mechanisms tailored for imaging sonar. Standard VAEs can be inefficient because of large latent spaces, while generic codebook methods often collapse or fail to capture subtle patterns. We introduce \textit{Adaptive Codebook Compression (ACC)} to learn frequency-encoded latent representations that stabilize training and preserve small, concealed objects. We further propose \textit{Frequency-Aware Multiscale Segmentation (FAMS)} to separate low-frequency background structures from sparse high-frequency dynamics, suppressing artifacts in the process. Finally, a \textit{hedging training strategy} leverages low-pass proxy pairs to guide frequency-aware learning without requiring clean–noise supervision. Together, these components enable SCOPE to jointly compress and correct sonar video, providing efficient low-bitrate transmission while retaining the details needed for reliable underwater monitoring.

\begin{figure}
\centering
  \includegraphics[width=0.5\textwidth]{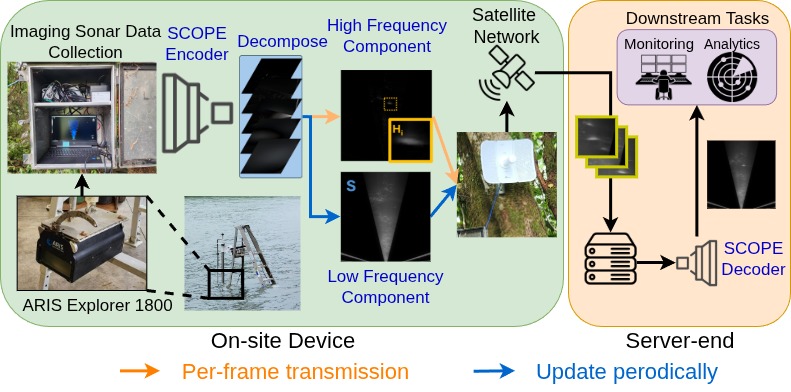}\
  \vspace{-0.7cm}
  \caption{SCOPE is a self-supervised framework for joint compression and artifact correction of imaging sonar streaming.} 
  \vspace{-0.6cm}
  \label{fig:introduction}
\end{figure}

SCOPE was evaluated on six months of ARIS sonar data collected from three Pacific Northwest rivers, providing a diverse and unbiased basis for testing. It was then deployed for an additional three months at the same sites (Fig.~\ref{fig:introduction}), supporting real-time salmon tracking, counting, and environmental monitoring. The framework achieved a structural similarity index (SSIM) of 0.77, representing a 40\% improvement over prior self-supervised denoising methods. It compressed sonar video to $\leq$ 0.0118 bits per pixel, reducing uplink bandwidth by more than 80\% while preserving image fidelity. These improvements also enhanced downstream detection and analysis. With 3.1 ms encoding on an embedded GPU and 97 ms full multi-layer decoding on the server, SCOPE operates in real time and demonstrates practical value under real-world deployment conditions.

The main contributions are summarized as follows:

\begin{itemize}
\item We develop and field-deploy SCOPE, a self-supervised framework that integrates compression and artifact correction to support real-time imaging sonar streaming under practical constraints.
\item We propose Adaptive Codebook Compression (ACC) (Sec.~\ref{sec:3.2}) and Frequency-Aware Multiscale Segmentation (FAMS) (Sec.~\ref{sec:3.3}) to jointly compress sonar data and suppress artifacts while preserving small, concealed signals.
\item We design a hedging training strategy (Sec.~\ref{sec:3.4}) that uses low-pass proxy pairs to guide frequency-aware learning without requiring clean–noise supervision.
\item Extensive experiments and in-situ deployments (Sec.~\ref{sec:4}) demonstrate that SCOPE achieves efficient low-bitrate transmission, improves image fidelity, and enhances downstream performance.
\end{itemize}

\section{Related Work}\label{sec:related work}
\subsection{Speckle-Affected Imaging Framework}

Change detection on speckle-contaminated images has been widely explored \cite{GONG2017212, 8713939, 10.1145/3666025.3699323}. Xu et al. \cite{10.1145/3666025.3699323} used Mixture of Gaussians (MOG) \cite{784637} with guided filtering \cite{6319316} to detect changes in sequential sonar frames, but pixel-wise modeling is computationally expensive, disrupts background consistency, and yields coarse, noisy outputs. Recent deep methods \cite{GONG2017212, 8713939} rely on pseudo labels from log-ratio images for CNN training, yet label noise introduces artifacts and hampers signal-preserving compression.

Despeckling remains challenging \cite{ZHAO2023103783,LI2025155,DALSASSO2024103940,9383788,10113743}. Li et al. \cite{LI2025155} combined change detection and attention but required dual-polarization inputs and labeled maps. Zhao et al. \cite{ZHAO2023103783} used contrastive and adversarial learning on unpaired data but struggled to generalize. Self-supervised methods show promise: MERLIN-Seg \cite{DALSASSO2024103940} still depends on labels, while Speckle2Void \cite{9383788}, based on Blind-Spot CNNs \cite{8954066}, oversimplifies speckle as uncorrelated noise.

\subsection{Learned Image Compression (LIC)}
Learned image compression (LIC) uses neural networks for efficient lossy compression, optimizing quality at low bitrates. Ballé et al. \cite{ballé2016,ballé2017,ballé2018} established the foundation for neural LIC. Recent methods include VAE-based models~\cite{10.1145/3664647.3681354,Jia_2024_CVPR,lu2024,Zhang_2024_CVPR,10.1145/3581783.3613799,10.1145/3664647.3681242,Duan_2024_CVPR}, diffusion models \cite{10.1145/3664647.3681336, Guo_2023}, and transformers ~\cite{Jiang_2023,10.1145/3581783.3611960,10.5555/3600270.3601221}. Guo et al. \cite{Guo_2023} introduced a diffusion model supporting flexible bitrate–quality trade-offs. VCT \cite{10.5555/3600270.3601221} tokenizes frames to capture spatio-temporal cues. ST-XCT \cite{10.1145/3581783.3611960} employs cross-covariance attention for improved coding. These designs improve performance but require heavy computation.

VAE-based CNNs remain dominant for latent modeling \cite{kingma2022}. GLC \cite{Jia_2024_CVPR} enhances perceptual quality via three-stage training in the VQ-VAE \cite{oord2018} latent space. STCM \cite{10.1145/3664647.3681242} captures temporal dynamics in remote sensing, while HybridFlow \cite{lu2024} reduces index map size without sacrificing quality. Bit Plane Slicing \cite{Zhang_2024_CVPR} combats posterior collapse \cite{lucas2019don} by emphasizing global features in hierarchical latents. Despite improved compression, these models involve multi-stage pipelines and high decoding cost, and often miss small or occluded signals—critical in sonar video analysis.

\subsection{Self-Supervised Image Denoising}
Self-supervised denoising has gained popularity due to scarce clean-noise pairs \cite{chihaoui2024, lehtinen2018, Zeng_2024_CVPR, chen2024, 8954066, laine2019, kim2024lan, huang2021}. Noise2Noise \cite{lehtinen2018} trains on noisy pairs, while Noise2Void \cite{8954066} and its extensions \cite{laine2019} use blind-spot networks (BSNs) to learn from single noisy inputs. However, these rely on ideal noise assumptions and degrade under real-world conditions. Neighbor2Neighbor \cite{huang2021} relaxes these assumptions via sub-sampling but still assumes IID noise, which fails for spatially correlated sonar noise \cite{10274511, rs16152815} (Fig.~\ref{fig:noise}).

To handle non-IID noise, newer approaches have emerged. Diff-Unmix \cite{Zeng_2024_CVPR} uses spectral unmixing with diffusion models, but is computationally intensive and assumes Gaussian noise. LAN \cite{kim2024lan} aligns real-world noise with pretrained priors, while MASH \cite{chihaoui2024} shuffles residuals, potentially introducing artifacts. AT-BSN \cite{chen2024} adjusts blind-spot size under weaker correlation assumptions—still ineffective for sonar data, where spatially dependent noise often obscures fine signals. In unpublished preprint work, SAVeD \cite{stathatos2025} exploits temporal cues from multiple frames to enhance low-SNR videos without clean labels. SCOPE instead integrates compression and artifact correction for real-time, low-bitrate analytics under deployment constraints.

\section{Background and Motivations}
Sonar devices have become increasingly affordable, from under \$2,000 for handheld units~\cite{field2025} to \$25,000 for advanced imaging sonar systems like ARIS~\cite{joslin2019imaging}, which are cheaper and higher-resolution than traditional sonar. Improved 5G and satellite connectivity enables broader real-time applications in remote areas, such as offshore rescue, disaster warnings, and fishery management. In turbid, low-light underwater environments where optical sensors fail, sonar provides a reliable alternative for rescues and fishery conservation~\cite{sonarrescue,10.1145/3666025.3699323}, highlighting its social impact in emergency response and environmental monitoring.

\noindent \textbf{Challenges from wild environments.} 
Despite wider availability, deploying sonar in remote areas faces hurdles from limited infrastructure. Terrestrial networks cover only 15\% of the Earth~\cite{cellularcoverage}, while satellite links suffer from high latency (up to 600 ms) and unstable uplink bandwidth, especially in dense vegetation, mountains, or harsh weather~\cite{10.1145/3664647.3680785}. These constraints hinder real-time imaging sonar transmission. E.g., an ARIS Explorer 3000~\cite{aris3000} at 1.8MHz outputs 1280×800@15 FPS, a downsample from higher native resolution used in previous deployments that degrades performance, still requiring $\geq$10.7 Mbps after codecs, whereas typical satellite uplinks provide only 4.7 Mbps (47\% of the required bandwidth), making compression essential.

\begin{figure}
\centering
  \includegraphics[width=0.48\textwidth]{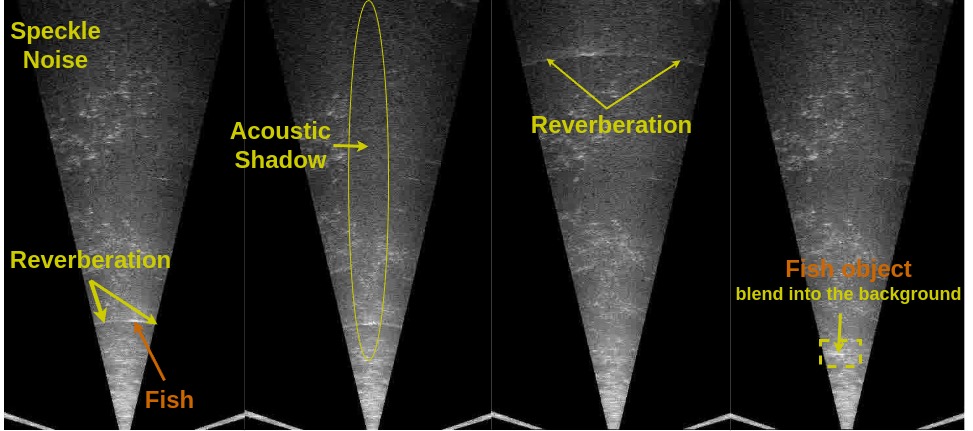}\
  \vspace{-0.6cm}
  \caption{Artifacts in data from underwater imaging sonar.}
  \vspace{-0.4cm}
  \label{fig:noise}
\end{figure}

\noindent \textbf{Artifacts of sonar data.}
Imaging sonar data from wild environments often contains artifacts such as speckle noise (high-frequency granular patterns), motion blur, acoustic shadows, and multipath reverberations (Fig.~\ref{fig:noise}) that obscure object boundaries, reduce contrast, lower perceptual quality, increase entropy, hinder compression, and lower analytic accuracy in tasks such as detection, tracking, and segmentation. Effective correction is essential for maintaining signal fidelity and reliable sonar imagery.

\begin{figure}
\centering
  \includegraphics[width=0.5\textwidth]{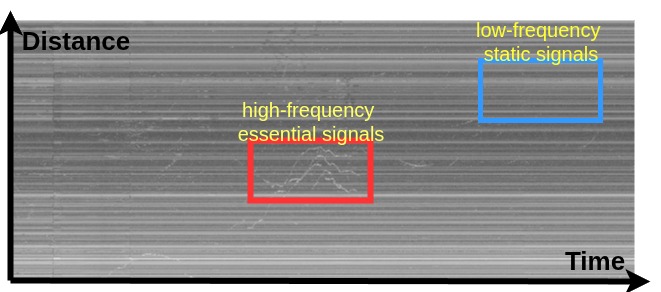}\
    \vspace{-0.6cm}
  \caption{Echograms show high-frequency signals as dynamic patterns and low-frequency signals as consistent static regions.
}
    \vspace{-0.5cm}
  \label{fig:echogram}
\end{figure}

\begin{figure*}
\centering
  \includegraphics[width=0.95\textwidth]{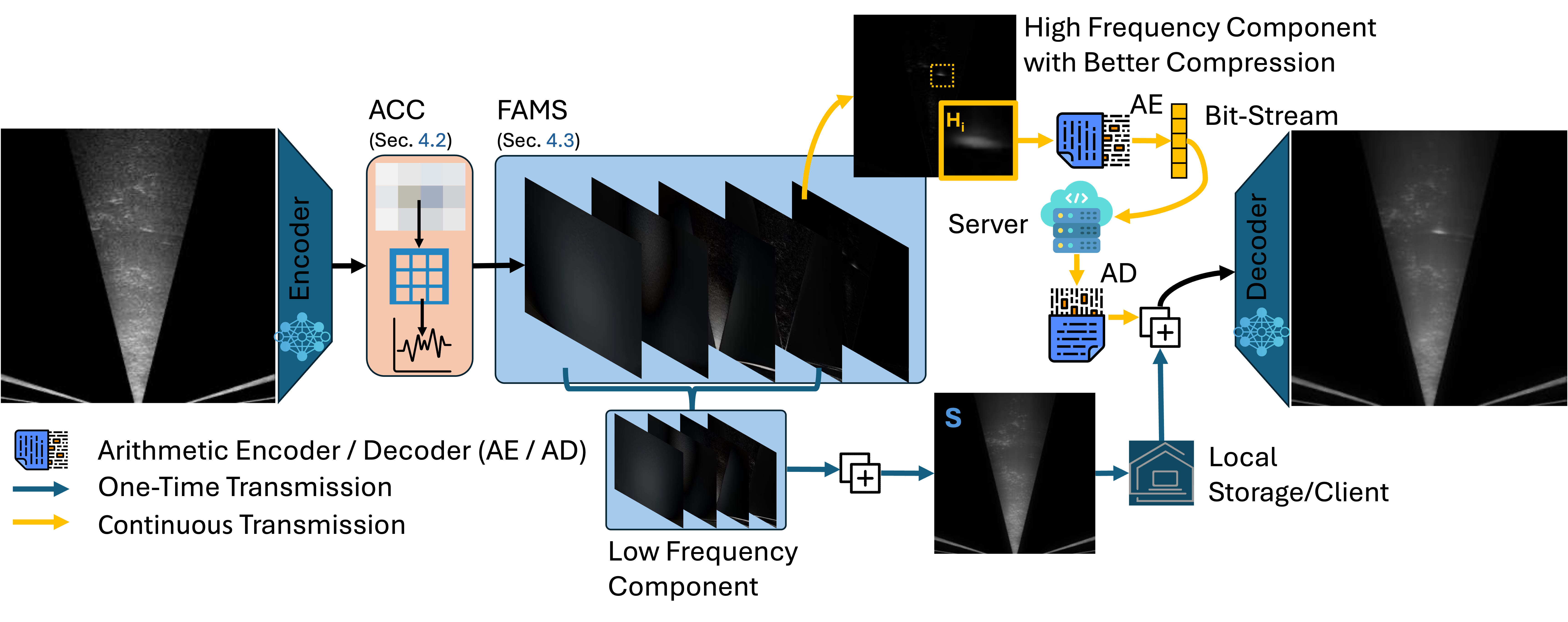}
      \vspace{-0.5cm}
  \caption{The overall underwater imaging sonar image compression and artifact correction architecture of SCOPE.}
      \vspace{-0.4cm}
  \label{fig:wacv26}
\end{figure*}

\section{Method}
We develop SCOPE, a self-supervised real-time imaging sonar streaming framework that jointly compresses data and corrects artifacts without requiring clean-noise pairs or synthetic data, and is deployed across three rivers (Fig.~\ref{fig:introduction}).

\subsection{SCOPE Overview}

SCOPE is motivated by temporal patterns observed in sonar echograms, where each vertical line represents a frame over time. As shown in Fig.~\ref{fig:echogram}, low-frequency static signals (blue box) dominate the scene, representing background and structured artifacts, while sparse high-frequency components (red box) capture dynamic targets such as marine life. The noise patterns in the echogram correspond to those in Fig.~\ref{fig:noise}, indicating strong spatial but weak temporal correlation. This observation suggests that frequency-domain segmentation is well-suited for denoising. Each frame \( I_i \) is modeled as:

\begin{equation} I_i = S + H_i + artifacts \end{equation}

As shown in Fig.~\ref{fig:wacv26}, a noisy input \( I_i \) is encoded into a latent representation \( \hat{I}_i \), which is processed by \textit{Adaptive Codebook Compression (ACC)} and \textit{Frequency-Aware Multiscale Segmentation (FAMS)}. ACC employs a VQ-VAE-based lossy LIC scheme to enhance representation learning of characteristic frequency components tailored for underwater imaging sonar data, producing \( k \) index maps for hierarchical sonar compression. FAMS performs soft frequency segmentation to isolate high-frequency components \( H_i \) (e.g., fish), generating \( k \) layers per frame by decoding the index maps \( k \) times (Fig.~\ref{fig:wacv26}, upper blue region). Compression is achieved by transmitting only \( H_i \) for each frame, while the static low-frequency background \( S \) is transmitted weekly. At the receiver, \( S \) and \( H_i \) are recombined to reconstruct a clean, artifact-free frame. Training is guided by a self-supervised hedging strategy (Sec.~\ref{sec:3.4}).

\begin{figure}
\centering
  \includegraphics[width=0.5\textwidth]{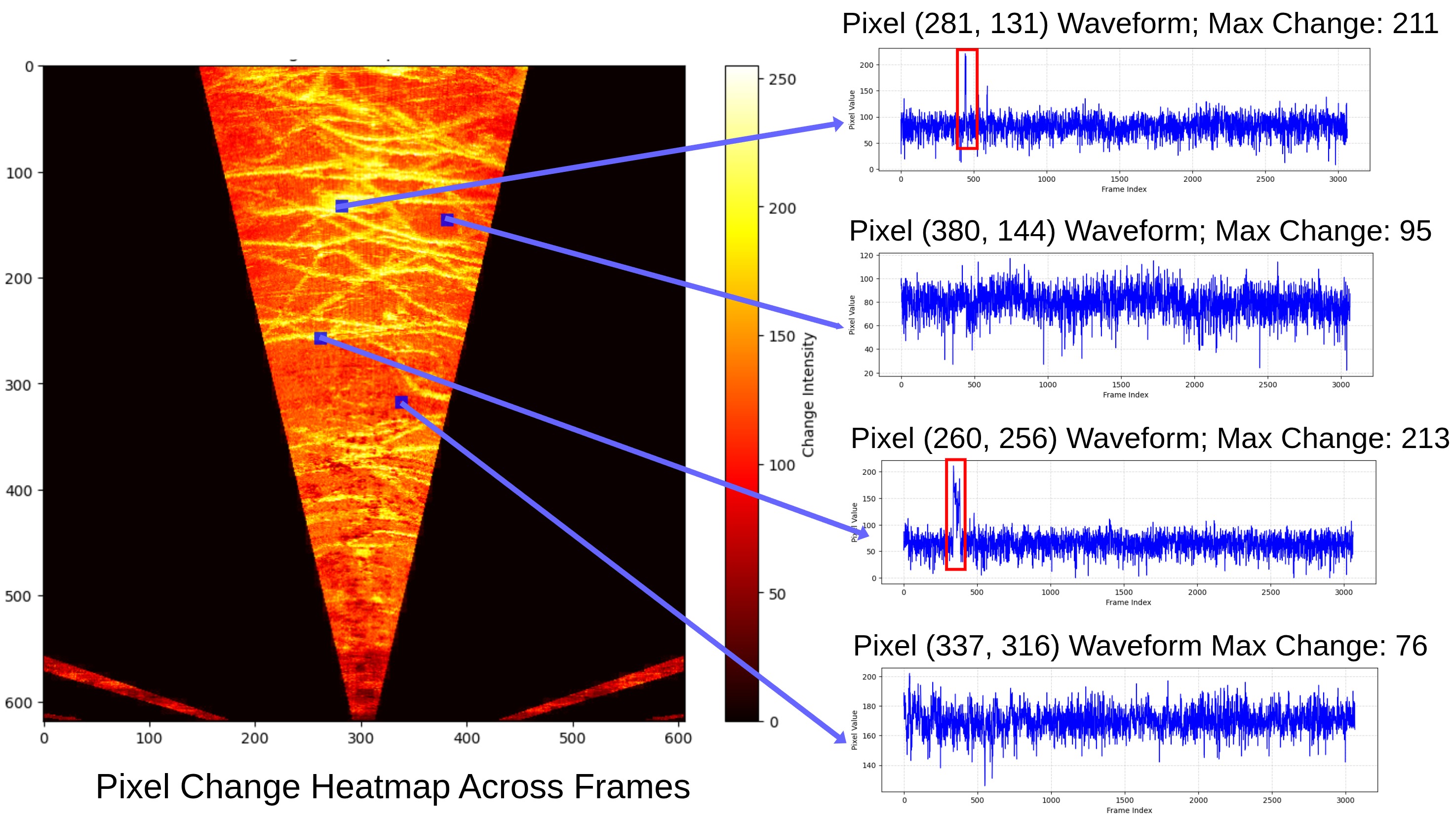}
  \vspace{-0.7cm}
  \caption{Pixel-wise temporal analysis of intensity fluctuations across imaging sonar frames.}
  \vspace{-0.7cm}
  \label{fig:Stat}
\end{figure}

\subsection{Adaptive Codebook Compression}\label{sec:3.2}
Learnable codebooks in VQVAE-based models~\cite{oord2018,esser2021,NEURIPS2024_9a24e284} face challenges such as large codebook sizes and posterior collapse~\cite{lucas2019don}, where an overly powerful decoder bypasses the latent space, limiting the model to a few mapped features. A rich codebook is essential for high-quality reconstruction, as its entries \( B_k \) are derived from latent features \( \hat{I}_i \). This work addresses two key questions: (1) How to preserve small, concealed objects blending into the background while avoiding collapse? (2) How to shrink the codebook to reduce bitrate?

Many signal representations rely on handcrafted transformations to extract frequency components, but such fixed bases often fail to capture diverse temporal dynamics. As shown in Fig.~\ref{fig:Stat}, sonar signals include distinct high-frequency essential components \( H_i \) (max change \( \geq 200 \)), low-frequency static components, and artifacts such as speckle noise, motion blur, and acoustic shadows (max change \( \leq 100 \)).

Our Adaptive Codebook Compression (ACC) departs from such fixed designs by introducing learnable codebook entries \( \mathbf{b}_k \), jointly optimized via the hedging training strategy (Sec.~\ref{sec:3.4}) to capture diverse frequency patterns. The codebook \( \mathbf{B} = [\mathbf{b}_1, \dots, \mathbf{b}_K] \) is trained to represent characteristic frequency components, enabling selective transmission of high-frequency \( H_i \) and low-frequency \( S \), yielding a compact and efficient representation.

\begin{figure}
\centering
    \includegraphics[width=0.45\textwidth]{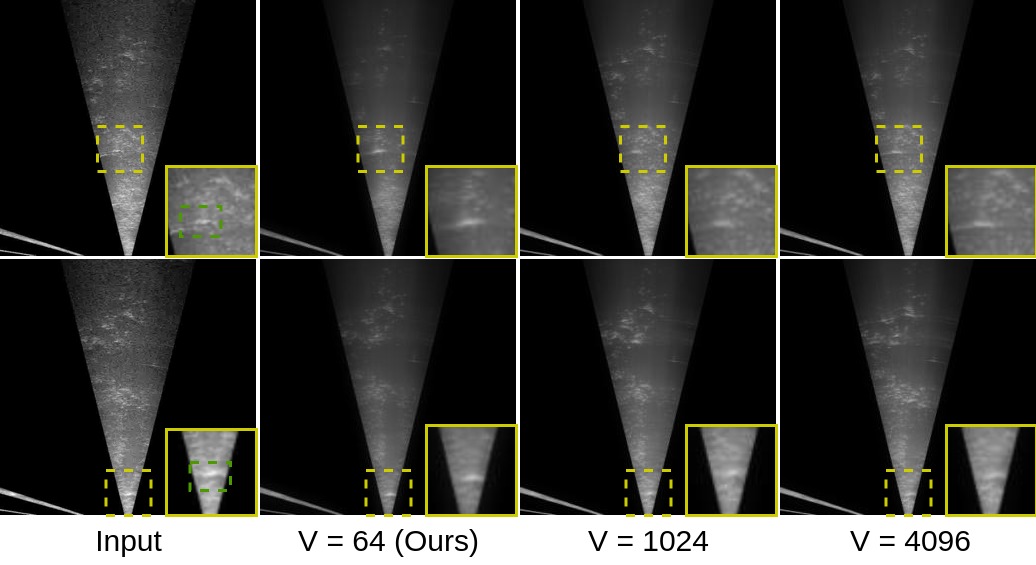}
    \vspace{-0.25cm}
    \caption{Quantitative comparisions on different codebook sizes $V \in \{64, 1024, 4096\}$. Smaller $V$ leads to better noise suppression and clearer separation of the objects.}
    \vspace{-0.5cm}

  \label{fig:as2}
\end{figure}

By exploiting limited frequency distributions and reducing entry size, ACC suppresses high-frequency artifacts with rapid fluctuations while preserving primary signal components (Fig.~\ref{fig:Stat}, red boxes), resulting in enhanced artifact suppression, improved compression efficiency, and better retention of small, background-blending objects (Fig.~\ref{fig:as2}).

\subsection{Frequency-Aware Multiscale Segmentation}\label{sec:3.3}
Fig.\ref{fig:Stat} shows a pixel-wise change heatmap across frames in real-world underwater imaging sonar video. Brighter areas (7.73\% of total), from white to yellow, indicate high-frequency object motions like fish and prawns (red box in Fig.\ref{fig:echogram}), while darker areas, from red to black, correspond to low-frequency static structures (blue box). The 2nd and 4th waveforms represent artifacts such as speckle noise, motion blur, and acoustic shadows. Comparing these waveforms reveals clear distinctions between high-frequency signals $H_i$, low-frequency static signals $S$, and artifacts.

FAMS decomposes signals into high-frequency temporal components (red box in Fig.~\ref{fig:Stat}) and low-frequency structural signals while suppressing rapidly fluctuating artifacts. This leverages image scaling, which shifts spatial frequency components. The transform of an image \(f(x,y)\) is given by:
\begin{equation}
F(u, v) = \int \int f(x, y) e^{-j2\pi (ux + vy)} \, dx \, dy
\end{equation}
Scaling by factor $s$ modifies the transform to:
\begin{equation}
F_s(u, v) = F\left( \frac{u}{s}, \frac{v}{s} \right)
\end{equation}

This scaling suppresses high-frequency components and enhances low-frequency ones. Unlike natural images, sonar images exhibit distinct frequency signatures (Fig.~\ref{fig:Stat}), making this operation effective for segmentation.
Learned entry vectors \( \mathbf{b}_k \) then transform these components, mapping spatial frequencies $F_s(u,v)$ to temporal domain vectors:

\begin{equation}
f = \text{lookup}(B, F_s)
\end{equation}

where $B$ is the codebook. The clean output is reconstructed by decoding the aggregated frequency components:

\begin{equation}f_{\text{denoised}}(x, y) = \text{Decoder}\left( \sum_k f_k \right)\end{equation}
\vspace{-0.25cm}

\begin{figure}
\centering
  \includegraphics[width=0.45\textwidth]{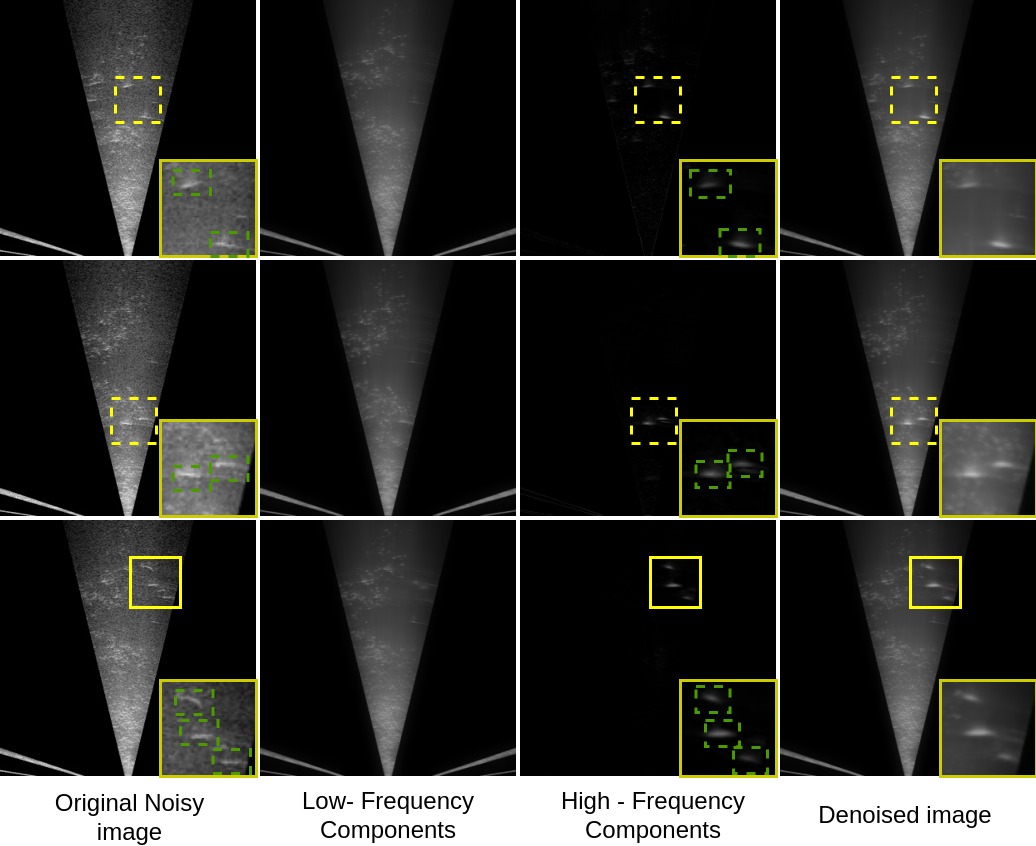}
  \vspace{-0.2cm}
  \caption{Effectiveness of our FAMS. The green boxes emphasize high-frequency dynamic signals (e.g., marine life).}
  \vspace{-1cm}
  \label{fig:recontrect}
\end{figure}

Fig.~\ref{fig:recontrect} demonstrates that FAMS, guided by the hedging strategy (Sec.~\ref{sec:3.4}), preserves high-frequency details \( H_i \) (highlighted by red boxes in Fig.~\ref{fig:Stat} and yellow boxes in Fig.~\ref{fig:recontrect}) in the final layer, while isolating low-frequency structural details with reduced artifacts in earlier layers. This enables effective frequency segmentation and artifact reduction, preserving small, concealed objects blending into the background. The method requires no sequential or temporal input, ensuring efficient and robust inference.

\subsection{Hedging Training Strategy}\label{sec:3.4}


To train SCOPE without clean-noise pairs, synthetic data, or noise assumptions, we propose a self-supervised hedging strategy that uses \textit{low-pass–filtered proxy pairs} to guide artifact correction and frequency-aware learning.

\begin{figure}
\centering
  \includegraphics[width=0.45\textwidth]{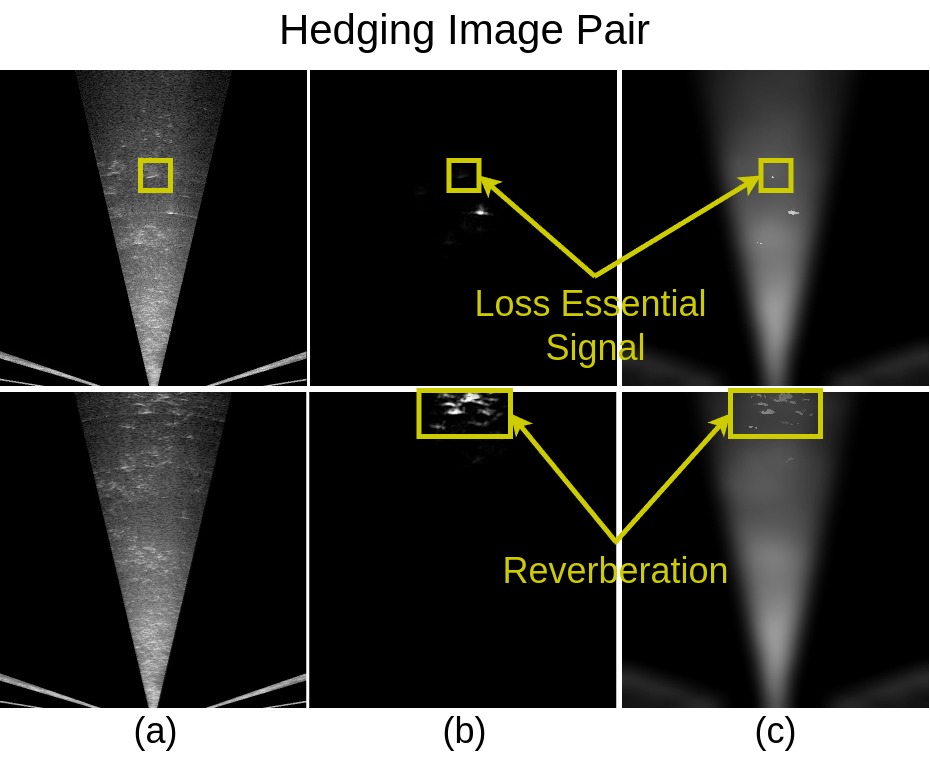}
  \vspace{-0.2cm}
  \caption{Low-pass proxy pair used in self-supervised learning, not directly applicable to denoising, reconstruction, or compression due to signal loss and reverberation.}
  \vspace{-0.7cm}
  \label{fig:compare_guided}
\end{figure}

\noindent \textbf{Low-pass proxy pair.}
As shown in Fig.~\ref{fig:compare_guided}(b)(c), we generate low-pass proxy pairs by first applying the Mixture of Gaussians (MOG) method for change detection to highlight dynamic elements (e.g., fish), followed by a guided filter acting as a low-pass filter. This process coarsely segments temporal frequency components, preserving low-frequency structures and edges while blurring static backgrounds, thereby emphasizing dynamic regions for segmentation guidance. This approach suits underwater imaging sonar data, where dynamic objects show higher temporal variance captured efficiently by MOG, and the guided filter preserves low-frequency spatial structures for robust feature learning. Combining temporal and spatial filtering provides a clear supervisory signal that emphasizes salient motion patterns while suppressing noise and static clutter, enabling effective learning of meaningful frequency representations.

However, low-pass proxy pairs are not directly applicable to artifact correction, reconstruction, or compression in streaming sonar. First, low-frequency regions are often distorted. Second, MOG fails to preserve high-frequency signals from slow or static objects (Fig.~\ref{fig:compare_guided}, top). Third, strong artifacts can introduce reverberation (Fig.~\ref{fig:compare_guided}, bottom). Lastly, generating low-pass proxy pairs requires sequential input and incurs high time complexity due to pixel-wise processing, making real-time edge deployment impractical.

\noindent \textbf{Hedging loss.} 
To enable both artifact correction and frequency-awareness, we compute the loss between the clean output, the noisy input, and the low-pass proxy pair:

\[
\mathcal{L} = \lambda_1 \mathcal{L}_{1}(\hat{I}, I_{noisy}) + \lambda_2 \mathcal{L}_{2}(\hat{I}, I_{hedging})
\]

Here, \( \hat{I} \) is the clean output, \( I_{noisy} \) is the noisy input, and \( I_{hedging} \) is the low-pass proxy pair. The first term, \( \mathcal{L}_{1}(\hat{I}, I_{noisy}) \), is a reconstruction loss that enforces structural consistency with the input to prevent over-smoothing and distortion. The second term, \( \mathcal{L}_{2}(\hat{I}, I_{hedging}) \), is a hedging loss that guides frequency-aware segmentation by leveraging the low-pass proxy pair, which highlights dynamic features and suppresses static background. This facilitates the removal of high-frequency artifacts and enables the decomposition of each frame into low-frequency structural components and high-frequency dynamic signals. 

The weighting factors \( \lambda_1 \) and \( \lambda_2 \) balance the contributions of the two loss terms, ensuring accurate restoration and effectively guiding frequency-aware segmentation.

\section{Experiment Results}
\label{sec:4}
\subsection{Experimental Settings}
\noindent \textbf{System setup and data collection.}
We deployed SCOPE (Fig.~\ref{fig:introduction}; more in supplementary) and a continuous underwater monitoring system in three Pacific Northwest rivers to enable real-time salmon tracking, counting, and monitoring, using the ARIS Explorer 1800 and ARIS Explorer 1200 Sonar systems. A Jetson ORIN Nano served as edge device, while data transmission and remote control were managed via Starlink. The underwater monitoring system ran for six months, producing 197,037 images and 1,346 videos with 454,941 annotated bounding boxes labeled by trained technicians under expert and biologist guidance.

\begin{figure*}
\centering
  \includegraphics[width=0.95\textwidth]{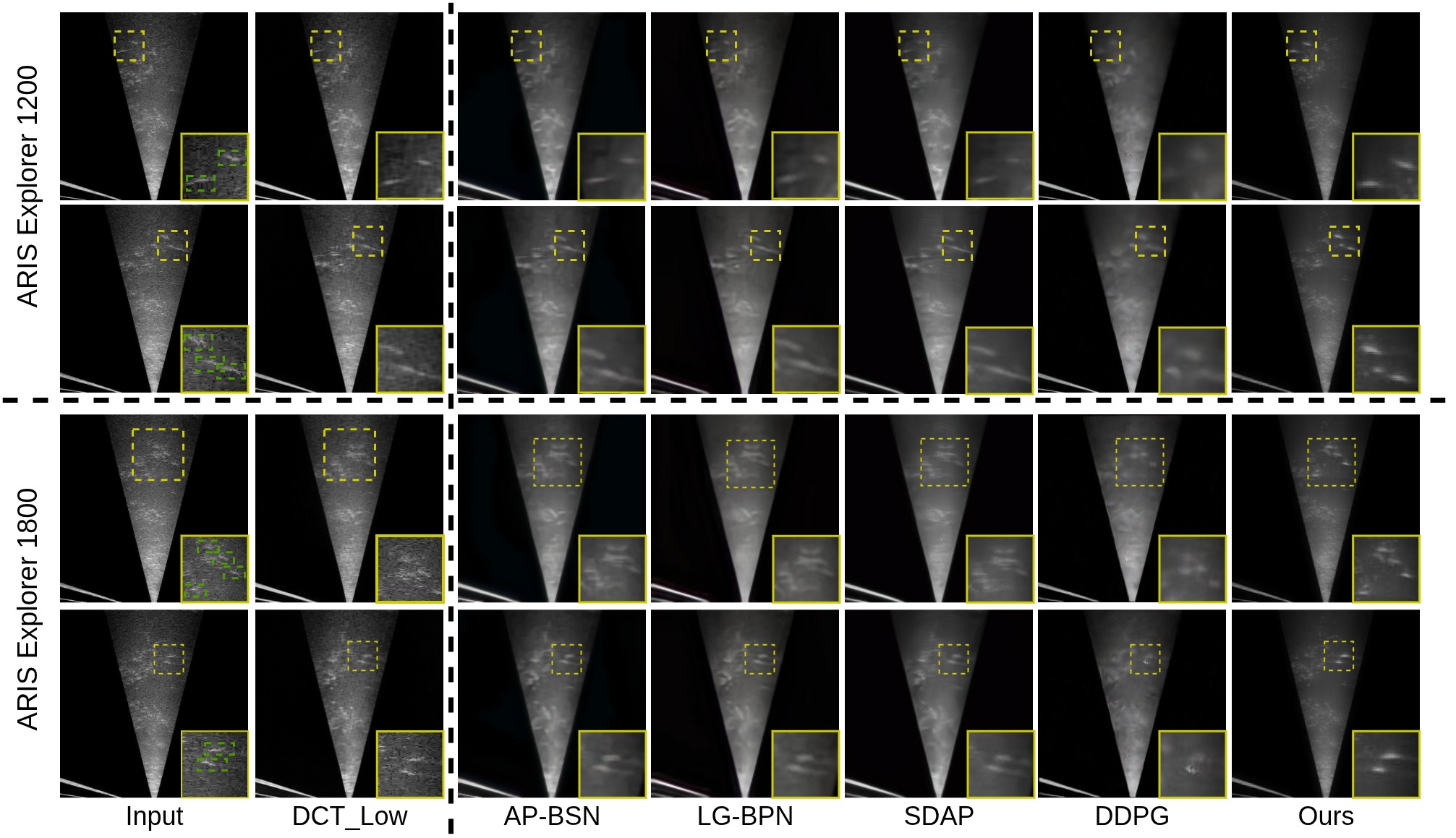}
  \vspace{-0.3cm}
  \caption{Visual comparison of self-supervised denoising methods on Underwater Imaging Sonar Dataset.}
  \vspace{-0.5cm}
  \label{fig:denoise_compare}
\end{figure*}

\noindent \textbf{Implementation details.}
Our model employs the vanilla VQ-VAE architecture \cite{oord2018} with a shared codebook size \( V = 64 \) as described in Section \ref{sec:3.2}. Instead of directly quantizing feature map values, the model is encouraged to learn frequency information through the hedging training strategy (Section \ref{sec:3.4}). The proposed FAMS uses a hierarchical architecture inspired by VAR \cite{NEURIPS2024_9a24e284}. Specifically, both ACC and FAMS are learning-based methods without preset rules or thresholds. They are implemented by generating layers through multiple guided VQ-VAE runs, each run producing a separate layer (Fig.~\ref{fig:wacv26}, blue region). For generating low-pass proxy pairs used in the loss computation (Section \ref{sec:3.4}), we apply a guided filter \cite{6319316}, an edge-preserving smoothing method that uses a guidance image for filtering. Specifically, two hedging images are created by applying the guided filter to the noisy image with MOG-based pixel-wise change detection as guidance, and vice versa. In the hedging loss, we set \(\lambda_1 = 0.6\) and \(\lambda_2 = 0.4\) to balance preserving structural similarity to the noisy input while effectively removing high-frequency artifacts and segmenting temporal frequency components into low-frequency (static background) and high-frequency (dynamic features).

\noindent \textbf{Inference time.}
On average, encoding takes 3.06 ms on the edge, while total decoding time for all layers averages 97 ms on the server. Since only the last layer ($\sim$ 26 ms) is typically needed, inference speed meets real-time requirements and supports efficient practical use. The effectiveness of the system in real-world scenarios is further demonstrated through downstream tasks detailed in Section \ref{sec:4.4}.

\noindent \textbf{Evaluation.}
As shown in Fig.~\ref{fig:introduction}, we carried out an in-field deployment of the proposed SCOPE across three rivers of the Pacific Northwest. We use SSIM and bpp as primary metrics since our model performs both compression and denoising. Since the output involves reconstruction and denoising, the original noisy input is not a suitable reference. Moreover, noise-free frames are hard to obtain due to device limitations and uncontrollable water conditions like flow speed and suspended particles. Considering these two factors, calculating metrics that require reference images such as PSNR is unsuitable. To further assess fidelity and denoising, we rely on downstream evaluation (Section \ref{sec:4.4}).

\subsection{Evaluation of Artifact Correction}\label{sec:4.1}
We target diverse artifact types rather than treating them as general noise, positioning denoising as a coarse correction step tailored to imaging sonar data. Our framework, SCOPE, is compared with state-of-the-art self-supervised denoising methods for real-world scenarios, including AP-BSN~\cite{9878719}, LG-BPN~\cite{10204077}, SDAP~\cite{10377346}, and DDPG~\cite{Garber_2024_CVPR}.

\begin{table}[t]
    \centering
    \caption{Denoising methods comparison on real-world sonar data.}
    \vspace{-0.2cm}
    \small
    \renewcommand{\arraystretch}{1.1} 
    \begin{adjustbox}{max width=0.48\textwidth}
    \begin{tabular}{c c c c}
        \toprule
          & SSIM \cite{1284395} $\uparrow$ & FSIM\cite{5705575} $\uparrow$ & BRISQUE \cite{6272356} $\downarrow$ \\
        \midrule \midrule
        AP-BSN \cite{9878719} & 0.5484 & 0.8882 & 76.0703  \\
        LG-BPN \cite{10204077} & 0.5440 & 0.8936 & 72.8588 \\
        SDAP \cite{10377346} & 0.5478 & 0.9158 & 78.2003 \\
        DDPG \cite{Garber_2024_CVPR} & 0.5434 & 0.9168 & 81.5875 \\
        \textbf{SCOPE(Ours)} & \textbf{0.7711} & \textbf{0.9208} & \textbf{70.8516}\\
        \bottomrule
    \end{tabular}
    \end{adjustbox}
    \vspace{-0.6cm}
    \label{tab:SSIM}
\end{table}

\noindent \textbf{Quantitative measure.}
Table~\ref{tab:SSIM} reports SSIM~\cite{1284395} scores to assess image quality relative to the noisy input. SSIM considers luminance, contrast, and structure, aligning better with human perception than PSNR. As a self-supervised framework, SCOPE achieves the highest SSIM (0.7711), significantly outperforming AP-BSN (0.5484), LG-BPN (0.5440), SDAP (0.5478), and DDPG (0.5434), demonstrating superior structural preservation and visual quality.

\noindent \textbf{Qualitative measure.}
Fig.~\ref{fig:denoise_compare} present qualitative comparisons in the image and frequency domains. DCT\_Low is reconstructed by applying inverse DCT to the 60×60 low-frequency region of the original 600×600 DCT from the noisy input, highlighting the importance of low-frequency information (further discussed in Sec.~\ref{sec:4.4}). As shown in Fig.~\ref{fig:denoise_compare}, SCOPE preserves fine details and more effectively recovers small, concealed objects blended into the background, which is critical for underwater sonar. Moreover, SCOPE suppresses high-frequency artifacts while retaining informative low-frequency signals (see Suppl.: Evaluation).



\subsection{Evaluation of Compression}
SCOPE is compared with existing VQ-VAE-based image compression methods, including HybridFlow~\cite{lu2024}, STCM~\cite{10.1145/3664647.3681242}, MLIC+~\cite{Jiang_2023}, and Continual-Compression \cite{Duan_2024_CVPR}. HybridFlow reduces bitrate via a masked codebook, while STCM and MLIC+ focus on entropy optimization. We evaluate SCOPE by comparing its bits per pixel (bpp) against the lowest bpp achieved by these methods. Unlike standard compression benchmarks, our model jointly performs reconstruction and denoising, making PSNR relative to the original image inapplicable.


\begin{table}[t]
    \centering
    \caption{Compression methods comparison on imaging sonar data.}
    \vspace{-0.2cm}
    \normalsize
    \renewcommand{\arraystretch}{1.1} 
    \begin{adjustbox}{max width=0.48\textwidth}
    \begin{tabular}{c c c c}
        \toprule
          & BPP $\downarrow$ & SSIM \cite{1284395} $\uparrow$ & BRISQUE \cite{6272356} $\downarrow$ \\
        \midrule \midrule
        \makecell{H.264 / H.265 / AV1\\(CRF:25)} & 0.7611 / 0.7396 / 0.6421  & - & -\\
        \midrule
        HybridFlow \cite{lu2024} & 0.025 & - & -  \\
        STCM \cite{10.1145/3664647.3681242} & 0.08 & - & - \\
        MLIC+ \cite{Jiang_2023} & 0.0975 & 0.7337 & 69.6311 \\
        Continual-Compression \cite{Duan_2024_CVPR} & 0.05 & 0.7781 & 75.2767 \\
        \midrule
        \makecell{\textbf{SCOPE(Ours)}\\(measured without entropy coding)} & $\leq$ \textbf{0.0118} & 0.7711 & 70.8516\\
        \bottomrule
    \end{tabular}
    \end{adjustbox}
    \vspace{-0.6cm}
    \label{tab:bpp}
\end{table}

Table~\ref{tab:bpp} reports the bpp performance. SCOPE achieves the lowest bpp (0.0118), measured without entropy coding. Compared with traditional codecs (H.264/H.265/AV1), it reduces bpp significantly from 0.64 to 0.0118. Since SCOPE outputs are predominantly dark, traditional codecs could further reduce redundancy, while SCOPE alone already cuts bitrate by at least 50\% compared to HybridFlow and over 87\% compared to MLIC+. This significant reduction demonstrates SCOPE’s efficiency in preserving critical features while minimizing bandwidth, making it highly suitable for resource-constrained streaming scenarios.

\subsection{Downstream task evaluation}\label{sec:4.4}
Image compression and denoising aim to reduce transmission cost and enhance image quality for perception and downstream applications, such as object detection. However, lossy compression may negatively impact downstream performance. To assess fidelity, we use detection on the original image as a baseline and compare it with compression methods, denoising methods and DCT\_Low (Sec.~\ref{sec:4.1}).

Using an 80/20 training/validation split, Table~\ref{tab:Downstream_task} shows that SCOPE achieves the highest AP50 (0.62947) and AP50-95 (0.22232), along with the best precision (0.75391), recall (0.57675), and lowest val/box loss (2.24997). DCT\_Low performs similarly to raw images, emphasizing the importance of preserving low-frequency components, while high-frequency details are less critical. Other denoising models perform poorly on sonar data, highlighting their limitations underwater. Under strict bandwidth constraints ($\sim$4.7 Mbps uplinks), SCOPE outperforms other compression approaches, achieving a 5\% improvement with lower bpp.

\begin{table}[t]
    \caption{Downstream object detection results on underwater sonar comparing raw source, compression, and denoising methods.}
    \vspace{-0.2cm}
    \centering
    \small
    \renewcommand{\arraystretch}{1.1} 
    \begin{adjustbox}{max width=0.48\textwidth}
    \begin{tabular}{c c c c c c}
        \toprule
        Method & AP50 & AP50-95 & Precision & Recall & box loss \\
        \midrule \midrule
        Raw Image & 0.6199 & 0.21521 & 0.73917 & 0.5669 & 2.26966 \\
        DCT\_Low & 0.61761 & 0.21476 & 0.73045 & 0.57402 & 2.27095 \\
        \midrule
        MLIC+ \cite{Jiang_2023}  & 0.57479 & 0.19126 & 0.69165 & 0.54103 & 2.33566 \\ 
        Continual-Compression \cite{Duan_2024_CVPR} & 0.57852 & 0.19575 & 0.70124 & 0.53802 & 2.32256 \\
        \midrule
        AP-BSN \cite{9878719} & 0.61382 & 0.21362 & 0.74366 & 0.56556 & 2.26588 \\
        LG-BPN \cite{10204077} & 0.61065 & 0.21177 & 0.73521 & 0.56488 & 2.28128 \\
        SDAP \cite{10377346} & 0.58852 & 0.20141 & 0.71599 & 0.54662 & 2.31268 \\
        \midrule
        \textbf{SCOPE(Ours)} & \textbf{0.62947} & \textbf{0.22232} & \textbf{0.75391} & \textbf{0.57675} & \textbf{2.24997} \\
        \bottomrule
    \end{tabular}
    \end{adjustbox}
    \vspace{-0.6cm}
    \label{tab:Downstream_task}
\end{table}

\subsection{Ablation Study}

\noindent \textbf{Adaptive Codebook Compression (ACC).}
We experiment with different codebook sizes $V \in \{64, 1024, 4096\}$ to show that reducing the codebook size not only achieves extremely low bitrates, but also encourages the codebook entries $\mathbf{b}_k$ to learn frequency components from temporal and spatial structures, rather than directly mapping feature values. As shown in Fig.~\ref{fig:as2}, smaller $V$ leads to better artifacts suppression and clearer separation of the objects.

\noindent \textbf{Frequency-Aware Multiscale Segmentation (FAMS).}
The effect of proposed FAMS is illustrated in Figs.~\ref{fig:wacv26} and \ref{fig:recontrect}. It suppresses high-frequency artifacts while segmenting low-frequency background structures and high-frequency object signals (e.g., fish). This method extends beyond imaging sonar, showing potential applicability in imaging domains such as MRI and remote sensing. In MRI, it could contribute to improved separation of tissue types. In remote sensing, it could contribute to more accurate change detection and geospatial monitoring.

\begin{figure}
  \includegraphics[width=0.45\textwidth]{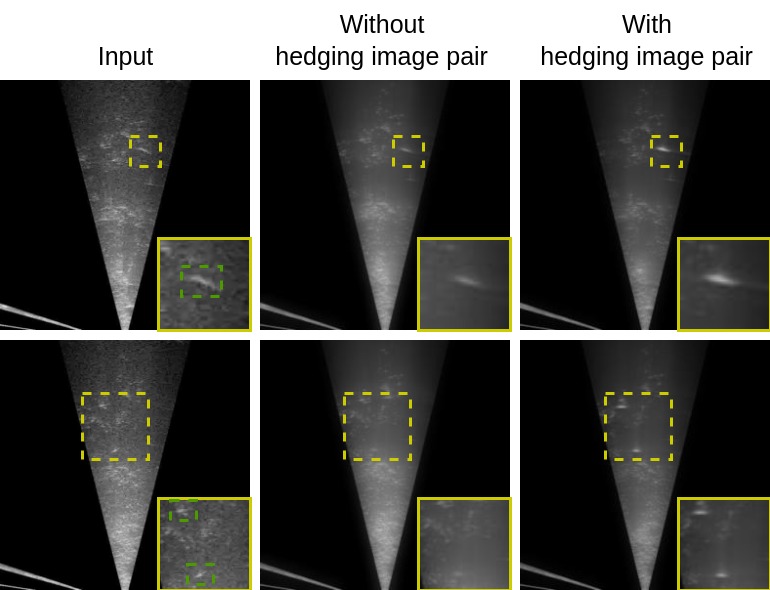}
  \vspace{-0.2cm}
  \caption{Effectiveness of our Hedging Training Strategy.}
  \vspace{-0.5cm}
  \label{fig:as1}
\end{figure}

\noindent \textbf{Hedging Training Strategy.}
We evaluate training with and without low-pass proxy pairs combined with different loss functions. Fig.~\ref{fig:as1} shows that the hedging strategy improves results and visual perception. Without hedging, small, concealed objects tend to blend into the background, causing erosion (row 1) and distortion (row 2) as seen in Fig.~\ref{fig:as1}.

\section{Conclusion}
In this paper, we presented a self-supervised SCOPE framework for underwater imaging sonar compression and artifact correction. By integrating ACC and FAMS with hedging training, it suppressed artifacts while preserving critical high-frequency signals. Experiments showed superior compression ($\leq0.0118$ bpp) suitable for 4.7 Mbps uplinks and artifact correction (SSIM: 0.7711), which improved downstream task accuracy. With 3.1 ms encoding and 97 ms total decoding for all layers, SCOPE enables real-time processing and has been deployed at three Pacific Northwest river sites for salmon and environmental monitoring, enabling practical sonar streaming. The method also shows promise for other imaging fields such as MRI and remote sensing.
\vspace{-0.1cm}
\section*{Acknowledgment}
\vspace{-0.1cm}
This research is supported by an NSERC Discovery Grant, a British Columbia Salmon Recovery and Innovation Fund (BCSRIF\_2022\_401), and a Mitacs Accelerate Cluster Grant. It also received additional funding from Experiment.com. Chi Xu's work is supported by an NSERC Canada Graduate Scholarship–Doctoral (CGS-D) and a Mitacs Globalink Research Award. We are grateful for the trust and collaboration of the Pacific Salmon Foundation, Wild Salmon Center, the Heiltsuk, Haida, Kitasoo Xai’xais, Taku River Tlingit, and Gitga’at First Nations, as well as the Skeena Fisheries Commission and Gitanyow Fisheries Authority for their ongoing partnership in this work.

{
    \small
    \bibliographystyle{IEEEtran}
    \bibliography{reference}
}

\end{document}